\documentclass{JFM-FLM_Au}
\usepackage{siunitx}
\usepackage[utf8]{inputenc}
\DeclareUnicodeCharacter{2212}{-}

\lefttitle{G.-Y. Ding et al.}
\righttitle{Journal of Fluid Mechanics}

\title{Scale-invariance and characteristic length scale for the large-scale vortices in geostrophic convective turbulence with friction}

\author{Guang-Yu Ding\aff{1,2}, Tian-Yi Pei\aff{1}, Hang-Yu Zhu\aff{3}, \and Ke-Qing Xia\aff{1,4}}

\affiliation{

\aff{1}Center for Complex Flows and Soft Matter Research and Department of Mechanics and Aerospace Engineering, Southern University of Science and Technology, Shenzhen, China
\aff{2}Research Institute of Intelligent Complex Systems, Fudan University, Shanghai, China
\aff{3}Department of Mechanics, School of Aerospace Engineering, Huazhong University of Science and
Technology, Wuhan, China
\aff{4}Department of Physics, Southern University of Science and Technology, Shenzhen, China
}

\corresau{Guang-Yu Ding, \email{dinggy@fudan.edu.cn}; Ke-Qing Xia, \email{xiakq@sustech.edu.cn}}

\begin{document}
\maketitle

\begin{abstract}
	In geostrophic convective turbulence, large-scale vortices (LSVs) emerge from the upscale energy transfer, usually accompanied with the large-scale friction. Yet, the role of large-scale friction in rotating convection remains poorly quantified. Here we use direct numerical simulation to study rotating Rayleigh--B\'{e}nard convection with a linear friction $\alpha\mathbf{u}$. Based on the classical KLB theory of 2D turbulence, a prediction $L_\alpha \sim \alpha^{-3/2}$ for the size of LSVs can be obtained. However, this prediction shows an apparent disagreement with the observed scaling $R_{LSV} \sim \alpha^{-1/2}$ for the LSV radius $R_{LSV}$. Such discrepancy arises from the energy spectrum of the barotropic (2D) manifold, which follows $E_{2D}(k)\sim k^{-3}$ in the range of upscale energy transfer, rather than the canonical KLB prediction $E(k)\sim k^{-5/3}$. Although this $k^{-3}$ spectrum has been widely reported in various systems, a clear theoretical interpretation for its origin remains lacking. Based on the energy pathways of the barotropic manifold, we show that the inverse energy cascade is significantly nonlocal, characterising by the coupling between various intermediate scales and the cutoff scale (the smaller one between the system scale and the frictional scale). This coupling leads to a strain balance across scales and a scale-invariant coarse-grained vorticity, which naturally yields the observed $k^{-3}$ spectrum. This argument is supported by the observed scaling of mean circulation $\langle|\Gamma(r)|\rangle\sim r^2$. The $k^{-3}$ spectrum directly leads to the $\alpha^{-1/2}$ scaling of $R_{LSV}$. These results provide a physical interpretation of the commonly observed $k^{-3}$ scaling in condensation-dominated turbulence, and futher suggest that predictions of the LSV size based on classical $k^{-5/3}$ spectrum may lead to a potential misestimation in geophysical and astrophysical systems.
\end{abstract}

\begin{keywords}
Geostrophic turbulence, B\'{e}nard convection, Turbulence theory
\end{keywords}

\section{Introduction}
Geostrophic turbulence is a fundamental process in many geophysical and astrophysical systems, such as the Earth's inner core \citep{cheng_2015_gji}, the ocean \citep{gayen_2022_arf}, tropical cyclones \citep{montgomery_2017_arf} and the Jovian atmosphere \citep{mitchell_2021_rg}. Although the Coriolis force introduced by rotation does not appear explicitly in the kinetic-energy budget, it can profoundly influence nonlinear interactions and hence the transfer of energy across scales. In classical three-dimensional (3D) turbulence, kinetic energy predominantly cascades forward from large to small scales. In contrast, rapidly rotating flows can become quasi-two-dimensional (quasi-2D), which promotes the features features of the 2D turbulence, such as an upscale transfer of energy \citep{kraichnan_1967_pof,benzi_elepl_1996_vsltf,chertkov_2007_prl,boffetta_arfm_2012_tt,alexakis_2018_physrep}. Energy then condenses at large scales and lead to the emergence of coherent large-scale vortex structure, which has been studied in both experiments \citep{morize_2005_pof,ruppert-felsot_2005_pre,kelley_2011_pof,zhu_2025_jfm} and numerical simulations \citep{smith_1999_pof,petersen_2006_pof,chertkov_2007_prl,buzzicotti_2018_prf,seshasayanan_2018_jfm}.

Among the various forcing mechanisms, buoyancy-driven turbulence is particularly important because of its close relevance to many geophysical and astrophysical systems . Unlike idealised models with a well-defined forcing scale, buoyancy in convection injects energy over a broad range of scales, leading to rich multiscale dynamics \citep{lohse_2010_arf}. A canonical example is rotating Rayleigh--B\'{e}nard convection (RRBC): a fluid layer heated from below and cooled from above, rotating about a vertical axis \citep{kunnen_2021_jtrb,ecke_2023_arf}. The buoyancy forcing and rotating constraint are respectively characterised by Rayleigh number $Ra\equiv(\beta_T g\Delta H^3)/(\kappa\nu)$, and Ekman number $Ek\equiv \nu/(2\Omega H^2)$. Here $\beta_T$ is the thermal expansion coefficient, $g$ is the gravitational acceleration, $\Delta$ the imposed temperature difference between the bottom and top plates, $H$ the layer height, $\kappa$ the thermal diffusivity, $\nu$ the kinematic viscosity, and $\Omega$ the rotation rate. For sufficiently large buoyancy forcing, the properties of quasi-2D turbulence emerge in RRBC, including the inverse cascade and energy condensation \citep{rubio_2014_prl}. One of the most significant impacts of the dimensionality transition in this system is the formation of large-scale vortices (LSVs) \citep{rubio_2014_prl,favier_2014_pof,cai_2021_apj,aguirre_2020_prl,de_wit_2022_jfm,van_kan_2025_jfm}. An LSV is a domain-scale 2D manifold induced by energy condensation. Similar flow structures can also be found in 2D turbulence \citep{chertkov_2007_prl} and thin-layer turbulence \citep{van_kan_2019_jfm}.

In most of the previous RRBC studies, the system is horizontally periodic and has little or no explicit large-scale energy sink. Consequently, energy transfers upscale without a natural cutoff, and then tends to condense at the domain scale. Such a process differs from geophysical settings, where large-scale frictional dissipation is ubiquitous, such as the Ekman friction associated with the Ekman boundary layers \citep{plumley_2016_jfm}. This linear friction provides a route to balance the inverse transfer at large scales and to produce finite-size large-scale structures. In classical 2D turbulence, energy cascades inversely through the inertial range from the forcing scale to large scales. The Kraichnan--Leith--Batchelor (KLB) theory proposes a scaling relationship $E(k)\sim \epsilon_F^{2/3}k^{-5/3}$ for this process, where $E(k)$ is the velocity energy spectrum in Fourier space, $k$ the wavenumber, and $\epsilon_F$ the external energy injection rate acting at the forcing scale $k_F$ \citep{kraichnan_1967_pof,leith_1968_pof,batchelor_1969_pof}. This upscale transfer continues until it eventually reaches the characteristic wavenumber of friction, namely $k_\alpha$. According to the KLB theory, the energy balance for this process can be written as

\begin{equation}
\epsilon_{F}\sim\sum_{k_\alpha<k<k_F}\alpha E(k)\sim\alpha\int_{k_F}^{k_\alpha}\epsilon_F^{2/3}k^{-5/3}dk\sim\alpha\epsilon_F^{2/3}k_\alpha^{-2/3},
\end{equation}
where $\alpha$ is the linear friction coefficient. It further yields

\begin{equation}
	\label{equ:l_alpha_k53}
	k_\alpha\sim\alpha^{3/2}\epsilon_F^{-1/2},\text{ and }L_\alpha=1/k_\alpha\sim\alpha^{-3/2}\epsilon_F^{1/2}.
\end{equation}
Equation \eqref{equ:l_alpha_k53} is widely used to estimate the cutoff scale of Ekman friction in many astrophysical systems. For example, \citet{shi_pnas_2026} uses \eqref{equ:l_alpha_k53} to estimate the cutoff scale of Ekman friction in the polar vortex model. However, the quantitative examination of equation \eqref{equ:l_alpha_k53} has received comparatively little attention, especially in rotating thermal convection.

Partially motivated by the issue mentioned above, we examine how large-scale friction controls the characteristic size and energy of LSVs in geostrophic convective turbulence in this work. We use direct numerical simulations to investigate a rotating Rayleigh--B\'{e}nard convection system with the linear friction term as a controllable large-scale sink. This configuration allows the flow to reach steady-state statistics and enables quantitative friction-controlled scalings. The numerical configuration and diagnostic procedures are described in sections \ref{sec:num_method}--\ref{sec:stat_method}. In section \ref{sec:size_of_lsv}, we characterise the LSV morphology and provide quantitative results on the LSV size as $\alpha$ varies. Interestingly, once friction dominates over finite-domain effects, we observe a scaling different from equation \eqref{equ:l_alpha_k53}. To understand the observed $\alpha$-dependence of the LSV scale, we conduct a 2D/3D (barotropic/baroclinic) manifold decomposition and analyse the associated energy spectra and transfers in section \ref{sec:manifold_decomposition}. We show that the inverse-transfer regime exhibits a robust $k^{-3}$ spectrum. We then provide a physical interpretation on the spectrum scaling based on a strain-balance argument, and demonstrate that this scaling directly yields the $\alpha$-dependence of the LSV scale. Finally, the main findings are summarised in section \ref{sec:conclusion}.

\section{Method}
\subsection{Numerical setup}
\label{sec:num_method}
In this study, we conduct direct numerical simulations (DNS) of three-dimensional rotating Rayleigh--B\'{e}nard convection. Under the Oberbeck--Boussinesq approximation, the nondimensional governing equations read

\begin{equation}
\label{equ:momentum_equ}
	\frac{\partial \mathbf{u}}{\partial t} + (\mathbf{u} \cdot \nabla) \mathbf{u} = -\nabla p + \sqrt{\frac{Pr}{Ra}} \nabla^2 \mathbf{u} + \theta \hat{e}_z-\frac{1}{Ek}\sqrt{\frac{Pr}{Ra}}\hat{e}_z\times\mathbf{u}-\alpha\mathbf{u},
\end{equation}
\begin{equation}
\label{equ:temperature_equ}
	\frac{\partial \theta}{\partial t} + (\mathbf{u} \cdot \nabla) \theta = \frac{1}{\sqrt{Ra Pr}} \nabla^2 \theta,
\end{equation}
\begin{equation}
\label{equ:divergence_free}
	\nabla \cdot \mathbf{u} = 0,
\end{equation}
where $\mathbf{u}$ is the velocity vector, $p$ is the pressure and $\theta$ is the temperature fluctuation. Here $Pr\equiv \nu/\kappa$ is the Prandtl number of the working fluid. The last term in \eqref{equ:momentum_equ} is a linear large--scale friction with nondimensional coefficient $\alpha$, defined as the ratio of the free-fall time scale $\tau_{ff}=\sqrt{H/(\beta_T g\Delta)}$ to a frictional time scale $\tau_\alpha=1/\alpha^*$, where $\alpha^*$ is the dimensional friction coefficient. We use this linear friction as a controllable large--scale sink for upscale energy transfer, applied to all velocity components.

\begin{figure}
	\centerline{\includegraphics[width=\textwidth]{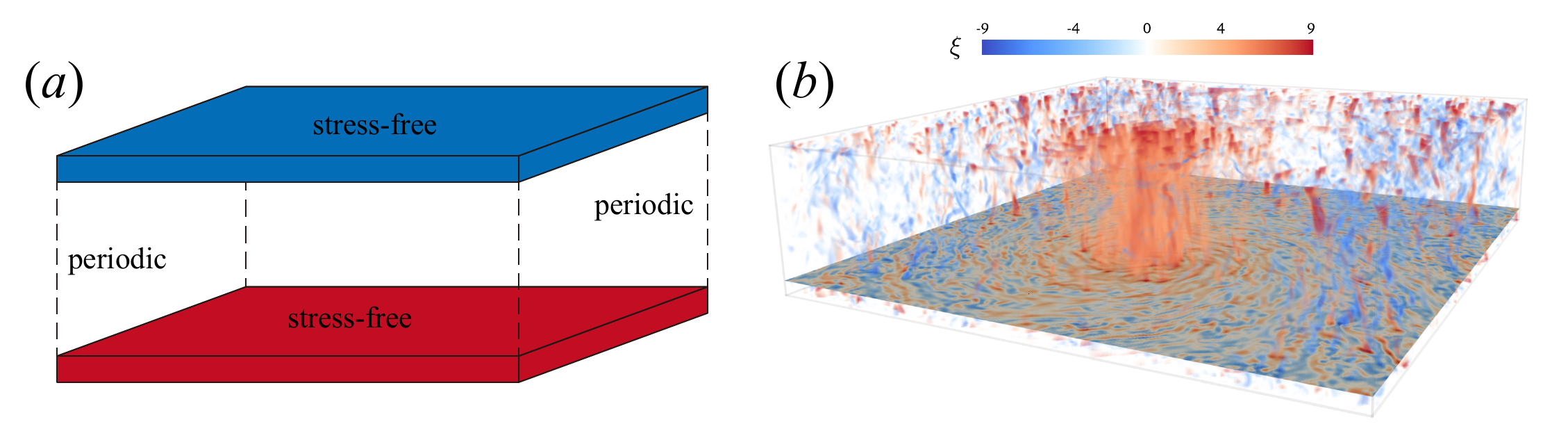}}
	\caption{(a) Sketch of the computational domain. (b) A snapshot of the volume-rendered vertical vorticity field for $\alpha=0$.}
	\label{fig:setup}
\end{figure}

In this study, we fix $Ra=10^8$, $Ek=3\times10^{-5}$ and $Pr=1$. The friction coefficient varies over $10^{-4}\leq\alpha\leq10^{-2}$, and we also run a frictionless baseline with $\alpha=0$. To provide a sufficiently large domain for LSV growth, we use an aspect ratio $L/H=4$, where $L$ is the horizontal domain width. Constant temperatures are imposed at the plates: $\theta_{z=1}=-0.5$ at the top and $\theta_{z=0}=0.5$ at the bottom. The lateral boundaries are periodic, and the top and bottom boundaries are stress-free. A schematic of the computational domain is shown in figure \ref{fig:setup}(a). The mesh resolution is $N_x\times N_y\times N_z=512\times 512\times 256$. The governing equations are solved with the well-tested CUPS code, based on a fourth-order finite-volume method \citep{kaczorowski_jfm_2013,kaczorowski_jfm_2014,chong_jcp_2018}. We use an identical initial field for all simulations, which is based on a previous simulation with $Ra=1\times10^8$, $Pr=8$, and similar boundary conditions as the current study.

\subsection{Details of LSV statistics}
\label{sec:stat_method}
\begin{figure}
	\centerline{\includegraphics[width=\textwidth]{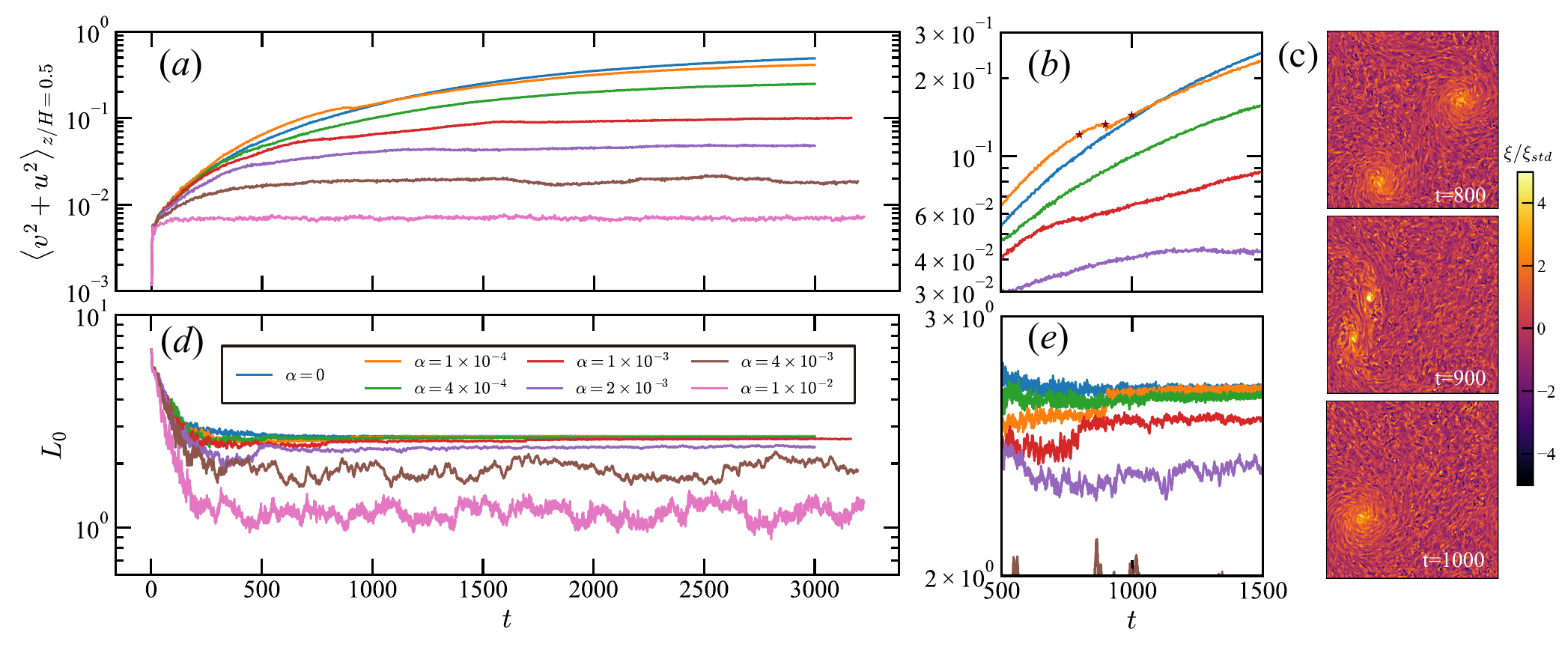}}
	\caption{Time traces of (a,b) horizontal kinetic energy and (d,e) the integral length scale $L_0$ at the midplane. (b,e) Zoomed-in views of the horizontal kinetic energy and $L_0$, respectively. (c) Vorticity distribution $\xi/\xi_{std}$ at the midplane in a LSV collision event for $\alpha=1\times10^{-4}$. The corresponding horizontal kinetic energy is indicated by the red stars in (b).}
	\label{fig:ke_l0}
\end{figure}

Owing to energy condensation, kinetic energy accumulate at large scales. For cases with weak or vanishing friction, there is no effective energy sink to dissipate this increasing large--scale energy. Thus, the system will take a very long time to reach a statistically steady state. Figure \ref{fig:ke_l0} presents the time traces of the mean horizontal kinetic energy $\langle u^2+v^2\rangle$ and the integral length scale $L_0$ evaluated at the horizontal midplane. The integral length scale $L_0$ is defined as

\begin{equation}
	L_0\equiv \frac{\int \hat{u}^2(k)/k dk}{\int \hat{u}^2(k) dk},
\end{equation}
where $\hat{u}(k)$ is the Fourier transform of the horizontal velocity component $u$. Both the mean kinetic energy and the integral scale provide useful indicators of LSV formation. Figure \ref{fig:ke_l0}(a) shows that the horizontal kinetic energy increases with time and eventually approaches a saturated level; in general, larger $\alpha$ yields earlier saturation and lower saturated energy. For the frictionless case ($\alpha=0$) and weak-friction cases (e.g.\ $\alpha=1\times10^{-4}$), the kinetic energy keeps rising over the entire simulation window, so a kinetic-energy steady state is not reached within the run time. By contrast, the integral length scale $L_0$ in figure \ref{fig:ke_l0}(d), which loosely tracks the LSV scale, behaves differently from the kinetic energy. For $\alpha\lesssim 1\times10^{-3}$, $L_0$ is nearly independent of $\alpha$, indicating that the large--scale vortices are limited by the domain size. The zoomed-in panels in figure \ref{fig:ke_l0}(b,e) reveal abrupt changes in the kinetic-energy and $L_0$ traces, associated with mergers of large-scale vortices. An example at $\alpha=1\times10^{-4}$ is shown in figure \ref{fig:ke_l0}(c), corresponding to the sudden drops of the orange curves in figure \ref{fig:ke_l0}(b,e). We regard the LSV structure as quasi-stable when $L_0$ fluctuates about a nearly constant mean and no further merger events are evident. On this criterion, stabilisation of the LSV structure is achieved much earlier than saturation of the large-scale kinetic energy. To balance statistical reliability and computational cost, we therefore accumulate statistics after $L_0$ has stabilised; full saturation of the kinetic energy is not required for every case. For weak or zero friction, where the kinetic energy still increases, we begin statistics once its growth rate has become sufficiently small. The corresponding start time is denoted by $t_{stat}$ and is listed in table \ref{tab:data_info} of the Appendix.

\begin{figure}
	\centerline{\includegraphics[width=0.8\textwidth]{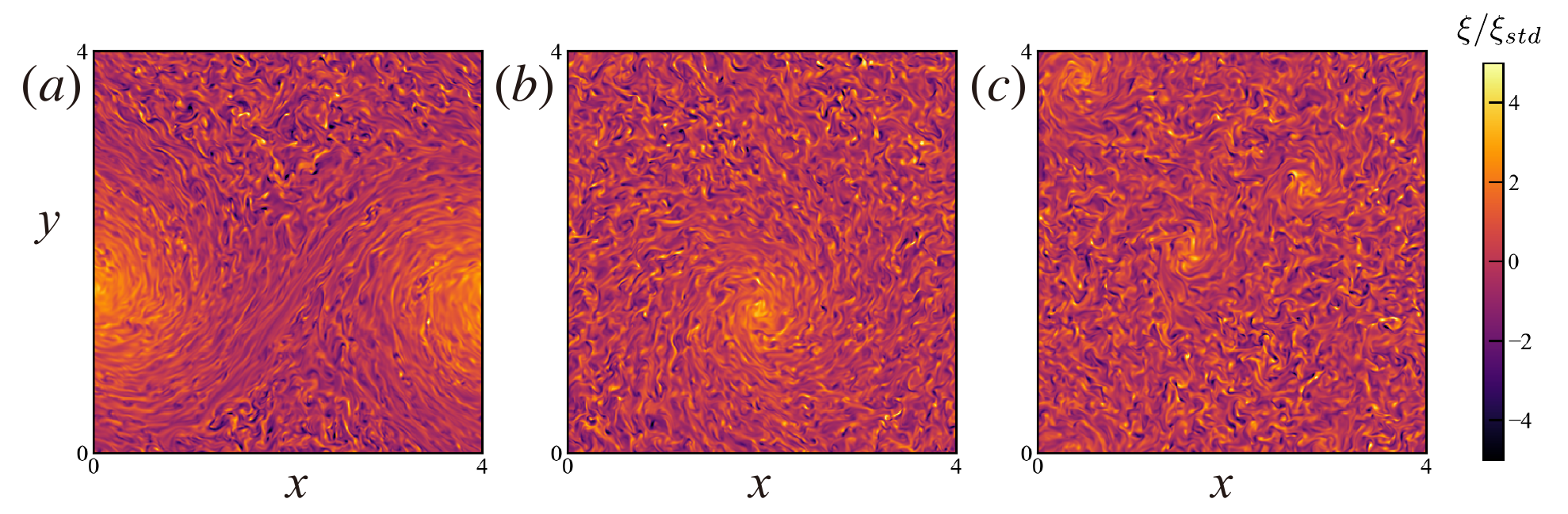}}
	\caption{Snapshots of the vertical vorticity field at the midplane for (a) $\alpha=0$, (b) $\alpha=1\times10^{-3}$, and (c) $\alpha=4\times10^{-3}$.}
	\label{fig:xi}
\end{figure}

Figure \ref{fig:xi} shows instantaneous snapshots of the normalised vertical vorticity $\xi/\xi_{std}$ at the midplane for several $\alpha$, where $\xi\equiv \partial v/\partial x - \partial u/\partial y$ and $\xi_{std}$ is the standard deviation of $\xi$. LSVs appear qualitatively as regions of strong vorticity, and their apparent size decreases as $\alpha$ increases. For a quantitative measurement of LSV scale, $L_0$ might seem a natural choice. We remark, however, that $L_0$ is not an accurate quantity to describe the size of LSV for detailed analysis: when background turbulence (no characteristic scale) is comparable with the LSV flow strength, as tends to occur at larger $\alpha$, $L_0$ can be biased and misrepresent the vortex size. In particular, a flow without a coherent LSV can still yield a finite $L_0$. For this reason, we therefore use $L_0$ mainly as a structural stability indicator. The LSV radius is measured from the radial profile of the ensemble-averaged LSV vorticity. Details of the measurement and related discussions are given in section \ref{sec:size_of_lsv}. Ensemble averaging requires robust identification of LSVs and their centres as well. We detect vortices from a filtered $Q$-field using the $Q$-criterion, and locate centres with a centre-of-mass procedure. Details of the identification strategy are given in the Appendix.

\section{Results}
\subsection{Flow structure and size of LSVs}
\label{sec:size_of_lsv}
\begin{figure}
	\centerline{\includegraphics[width=0.8\textwidth]{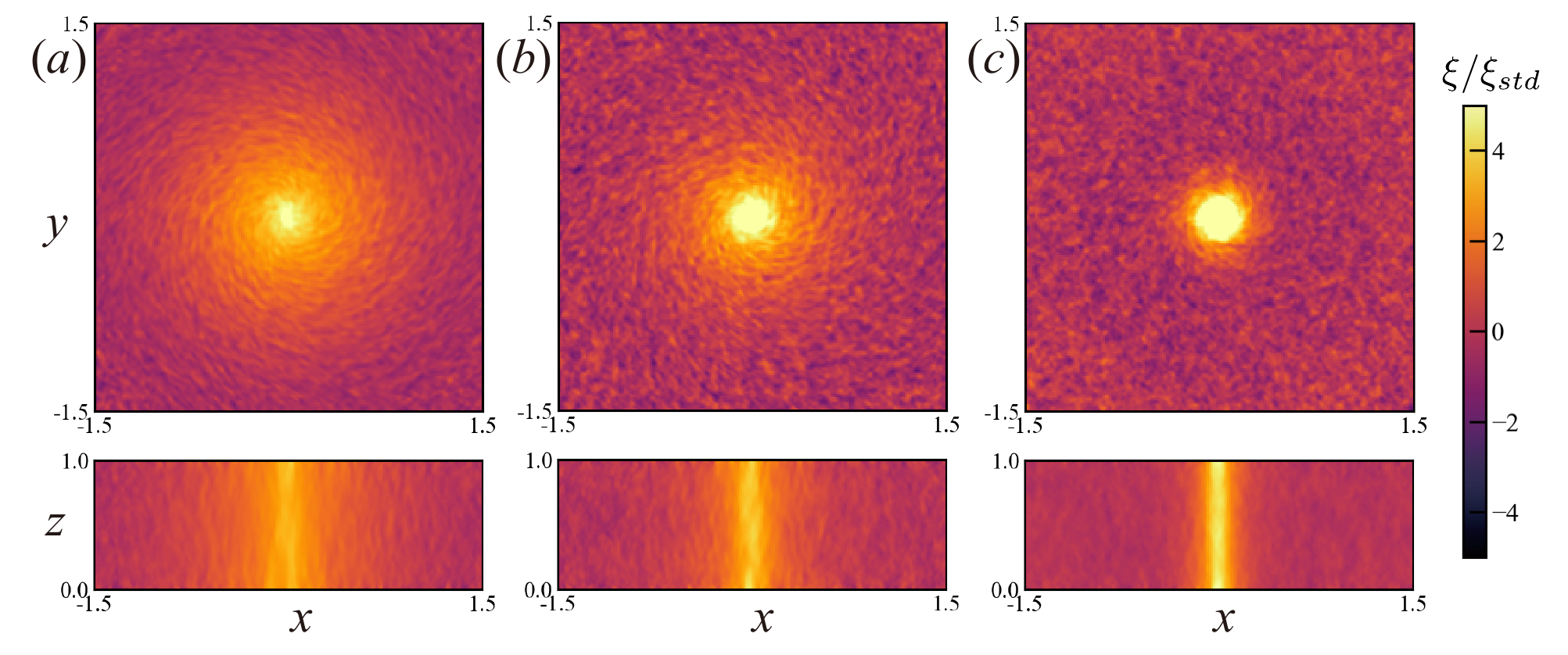}}
	\caption{Ensemble average of the vertical vorticity distribution at the horizontal midplane (upper panel) and vertical plane (lower panel) for (a) $\alpha=0$, (b) $\alpha=1\times10^{-3}$, and (c) $\alpha=4\times10^{-3}$.}
	\label{fig:xi_ensemble}
\end{figure}
Figure \ref{fig:xi_ensemble} presents the ensemble average of the normalized vertical vorticity $\xi/\xi_{std}$ at the midplane (upper panels) and vertical plane (lower panels) for different $\alpha$. The vorticity is roughly uniform in the vertical direction, consistent with the understanding of LSVs as a manifestation of the transition to quasi-2D turbulence \citep{rubio_2014_prl,favier_2019_jfm}. The mean vorticity of the whole volume should be zero, owing to the zero mean circulation in a periodic domain. However, all LSVs in this study have positive vorticity, while the background turbulent vorticity field is mainly negative. This asymmetry indicates that the cyclone--anticyclone symmetry is broken, consistent with previous findings \citep{favier_2014_pof}. A recent numerical study shows that the cyclone--anticyclone symmetry recovers if $Ek$ is further decreased \citep{van_kan_2025_jfm}. Nevertheless, in our broken-symmetry cases, we can straightforwardly ensemble the identified LSVs and compute the average.

\begin{figure}
	\centerline{\includegraphics[width=\textwidth]{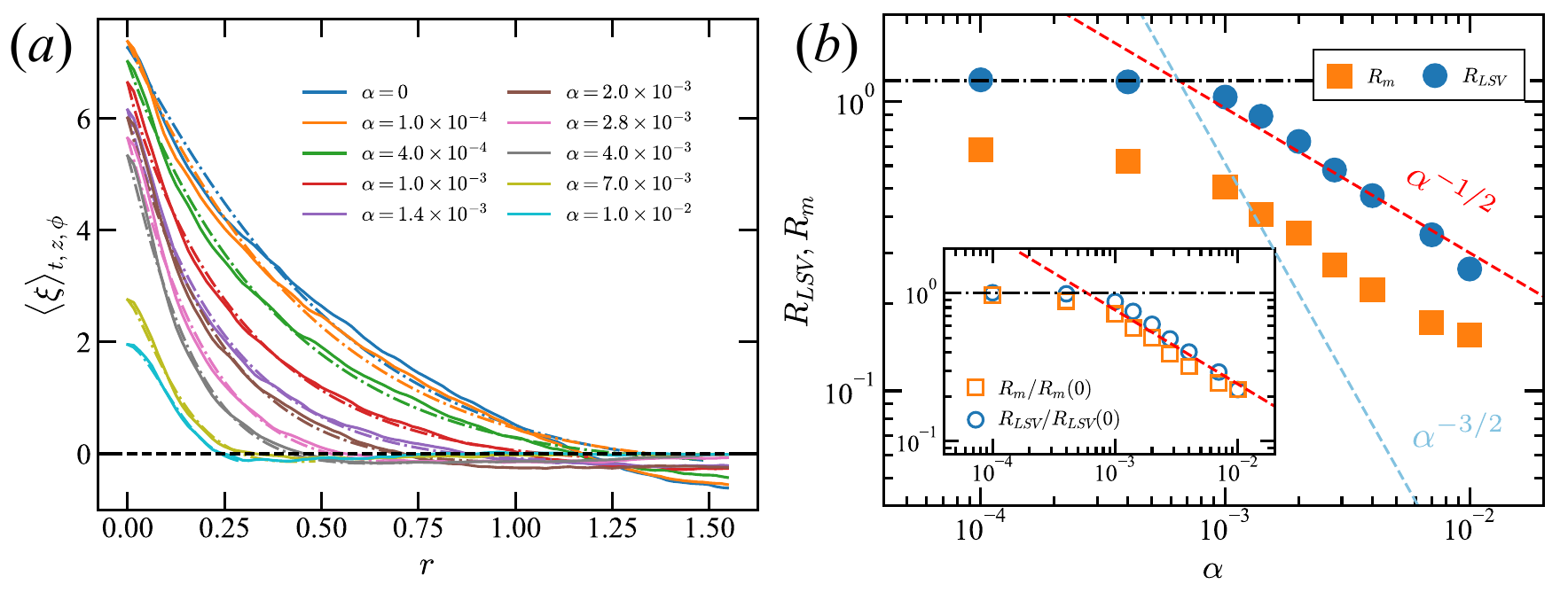}}
	\caption{(a) Radial distribution of the ensemble-averaged LSV vorticity $\langle\xi\rangle_{z,\phi}$. The dot-dashed curves correspond to the fitting using equation \eqref{equ:xi_radial}. (b) Scales of the LSV quantified by $R_{LSV}$ (blue circles) and $R_m$ (orange squares) as $\alpha$ changes, respectively. The black dot-dashed line denotes the result of the frictionless case $\alpha=0$, and the red and light blue dashed lines are guides to the eye for the scaling relations $\alpha^{-1/2}$ and $\alpha^{-3/2}$, respectively. The inset presents normalized plot for $R_m/R_m(0)$ and $R_{LSV}/R_{LSV}(0)$, where $R_m(0)$ and $R_{LSV}(0)$ correspond the values with $\alpha=0$.}
	\label{fig:rlsv}
\end{figure}

Based on the ensemble-averaged vorticity field, we can quantitatively define the size of LSVs. In figure \ref{fig:rlsv}(a), we present the radial distribution of the ensemble-averaged LSV vorticity. We use equation (A1) in \citet{couston_prr_2020} to describe the radial profile of the azimuthal velocity, which gives the distribution of LSV vorticity as follows

\begin{equation}
	\label{equ:xi_radial}
	\xi=\xi_0e^{-\frac{1}{\mu}\left(\frac{r}{R_m}\right)^\mu}\left[1-\frac{1}{2}\left(\frac{r}{R_m}\right)^\mu\right],
\end{equation}
where $\xi_0$ is the vorticity at $r=0$, $R_m$ corresponds to the radius with maximum azimuthal velocity, and $\mu$ is a tuning parameter. For $\mu=1$ and $R<R_m$, equation \eqref{equ:xi_radial} reduces to an exponential decay $\xi\approx \xi_0 e^{-r/R_m}$. We fit the radial vorticity profile of the LSVs using equation \eqref{equ:xi_radial}, and list the obtained fitting parameters in table \ref{tab:data_info}. The dot-dashed curves in figure \ref{fig:rlsv}(a) refer to the fitting. The vorticity profile is well described by equation \eqref{equ:xi_radial}. According to the physical meaning of $R_m$ in \eqref{equ:xi_radial}, it is natural to use $R_m$ to characterise the size of LSVs. For an alternative measure of the LSV scale, the maximum radius of the LSV can be defined as the position where the vorticity declines to zero. Thus, we define the radius of LSV as the zero-crossing radius of $\langle\xi\rangle_{t,z,\phi}$. In figure \ref{fig:rlsv}(b), we plot $R_m$ and $R_{LSV}$ against $\alpha$. The black dot-dashed line denotes the result of $R_{LSV}$ for the frictionless case $\alpha=0$. For weak or vanishing friction, the scale of LSV is constrained by the domain size and independent of $\alpha$. As $\alpha$ increases, the cutoff mechanism switches from the finite-size effect to the large-scale friction. In this case, we observe that $R_m$ and $R_{LSV}$ decrease with the friction coefficient $\alpha$. 

Once the friction becomes crucial, we surprisingly find that both $R_m$ and $R_{LSV}$ follow a scaling relationship $\alpha^{-1/2}$ (red dashed line), instead of the predicted scaling $\alpha^{-3/2}$ in equation \eqref{equ:l_alpha_k53} (blue dashed line). In the inset of figure \ref{fig:rlsv}(b), the normalised LSV scales exhibit a good agreement with $\alpha^{-1/2}$. On the other hand, the $\alpha^{-3/2}$ scaling severely deviates from the observed results, suggesting that equation \eqref{equ:l_alpha_k53} cannot accurately describe the cutoff scale for the friction in geostrophic convective turbulence. In order to explain such discrepancy and, more importantly, understand the physical interpretation for the observed $\alpha^{-1/2}$ scaling, we need to first re-examine how \eqref{equ:l_alpha_k53} is derived. The scaling \eqref{equ:l_alpha_k53} is built on the balance between the upscale energy transfer and large-scale friction. The exponent $-3/2$ relies on the scaling of energy spectrum $E(k)\sim k^{-5/3}$ in the inertial range of inverse energy cascade. However, there is in fact evidence that the energy spectrum follows an alternative scaling $E(k)\sim k^{-3}$ during energy condensation in, for example, 2D turbulence \citep{chertkov_2007_prl}, thin-layer turbulence experiments \citep{zhu_2024_jfm}, and rotating thermal convection, which is the same system as the current study \citep{rubio_2014_prl}. For this reason, the energy transfer is crucial to explain observed LSV scale in figure \ref{fig:rlsv}(b).

\subsection{Manifold decomposition and energy transfer}
\label{sec:manifold_decomposition}
In this section, we focus on the energy transfer and spectrum properties. To distinguish the flow dynamics pertaining to the LSV from the background turbulence, we perform a 2D--3D decomposition of the field. The 2D $\mathbf{u}_{2D}$ and 3D manifolds $\mathbf{u}_{3D}$ are respectively defined as $\mathbf{u}_{2D}\equiv\overline{\mathbf{u}}$, and $\mathbf{u}_{3D}\equiv \mathbf{u}-\mathbf{u}_{2D}$, where $\overline{*}$ denotes the average in the vertical direction. The 2D (3D) manifold is also denoted as the slow (fast) or barotropic (baroclinic) mode in the literature \citep{alexakis_2018_physrep,rubio_2014_prl}. The governing equation for the 2D (3D) manifold can be respectively written as 
\begin{equation}
	\label{equ:2d_manifold}
	\begin{split}
	\partial_t \mathbf{u}_{2D}+(\mathbf{u}_{2D}\cdot \nabla)\mathbf{u}_{2D}=&-\overline{\nabla p}+\sqrt{\frac{Pr}{Ra}}\nabla^2 \mathbf{u}_{2D}\\
	&-\frac{1}{Ek}\sqrt{\frac{Pr}{Ra}}\hat{e}_z\times \mathbf{u}_{2D}-\alpha \mathbf{u}_{2D}\\
	&-\overline{\mathbf{u}_{3D}\cdot\nabla \mathbf{u}_{3D}},
	\end{split}
\end{equation}
and 
\begin{equation}
	\label{equ:3d_manifold}
	\begin{split}
	\partial_t \mathbf{u}_{3D}+(\mathbf{u}_{3D}\cdot \nabla)\mathbf{u}_{3D}=&-(\nabla p-\overline{\nabla p})+\sqrt{\frac{Pr}{Ra}}\nabla^2 \mathbf{u}_{3D}\\
	&-\frac{1}{Ek}\sqrt{\frac{Pr}{Ra}}\hat{e}_z\times \mathbf{u}_{3D}-\alpha \mathbf{u}_{3D}
	+\alpha g\theta \hat{e}_z \\
	&+\overline{\mathbf{u}_{3D}\cdot\nabla \mathbf{u}_{3D}}-(\mathbf{u}_{3D}\cdot\nabla)\mathbf{u}_{2D}-(\mathbf{u}_{2D}\cdot\nabla)\mathbf{u}_{3D}.
	\end{split}
\end{equation}
The first rows of equations \eqref{equ:2d_manifold} and \eqref{equ:3d_manifold} correspond to the standard Navier--Stokes terms; the second rows describe the external forcing, including the Coriolis force and the large--scale friction; the third rows correspond to the manifold-to-manifold interactions. The dynamics of the LSV is given by the 2D manifold \eqref{equ:2d_manifold}. It is worth mentioning that buoyancy does not appear in equation \eqref{equ:2d_manifold}, meaning that the LSV is not sustained by buoyancy, but instead by the manifold-to-manifold interactions in the third row of equation \eqref{equ:2d_manifold}. 

\begin{figure}
	\centerline{\includegraphics[width=\textwidth]{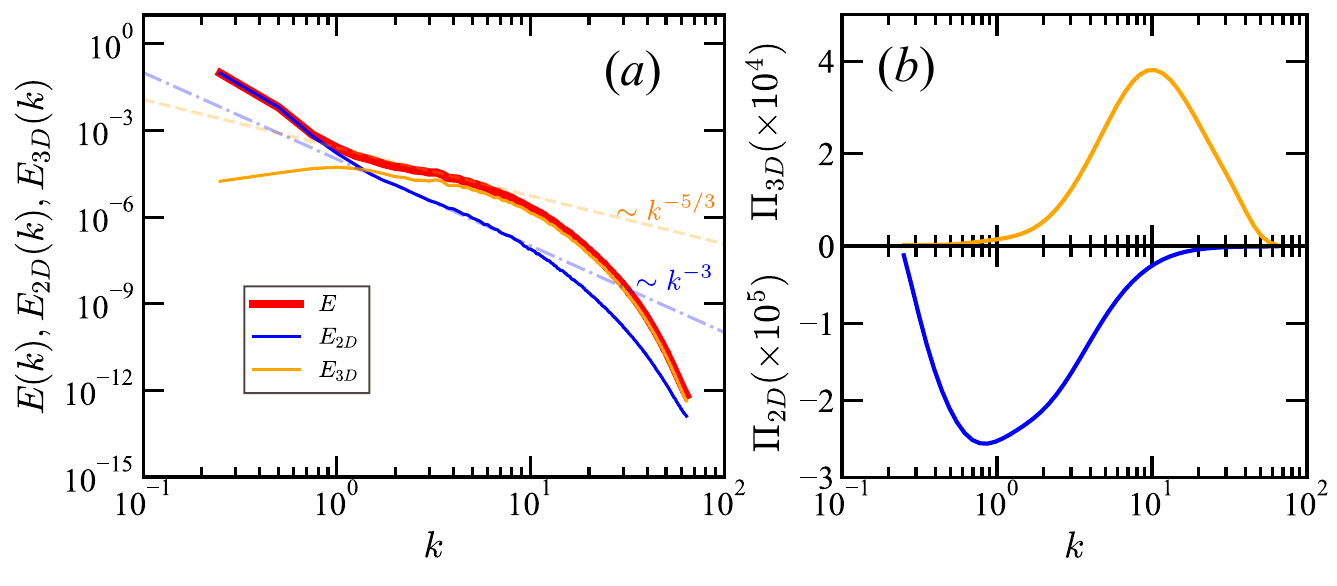}}
	\caption{(a) Energy spectrum for $\alpha=1\times10^{-3}$. The orange dashed line denotes $E(k)\sim k^{-5/3}$, while the blue dot-dashed line denotes $E(k)\sim k^{-3}$. (b) Energy flux for 3D manifold $\Pi_{3D}$ and 2D manifold $\Pi_{2D}$ respectively. A prefactor of $10^4$ and $10^5$ is applied to the flux for better visualization.}
	\label{fig:power_spectrum_single}
\end{figure}

\begin{figure}
	\centerline{\includegraphics[width=\textwidth]{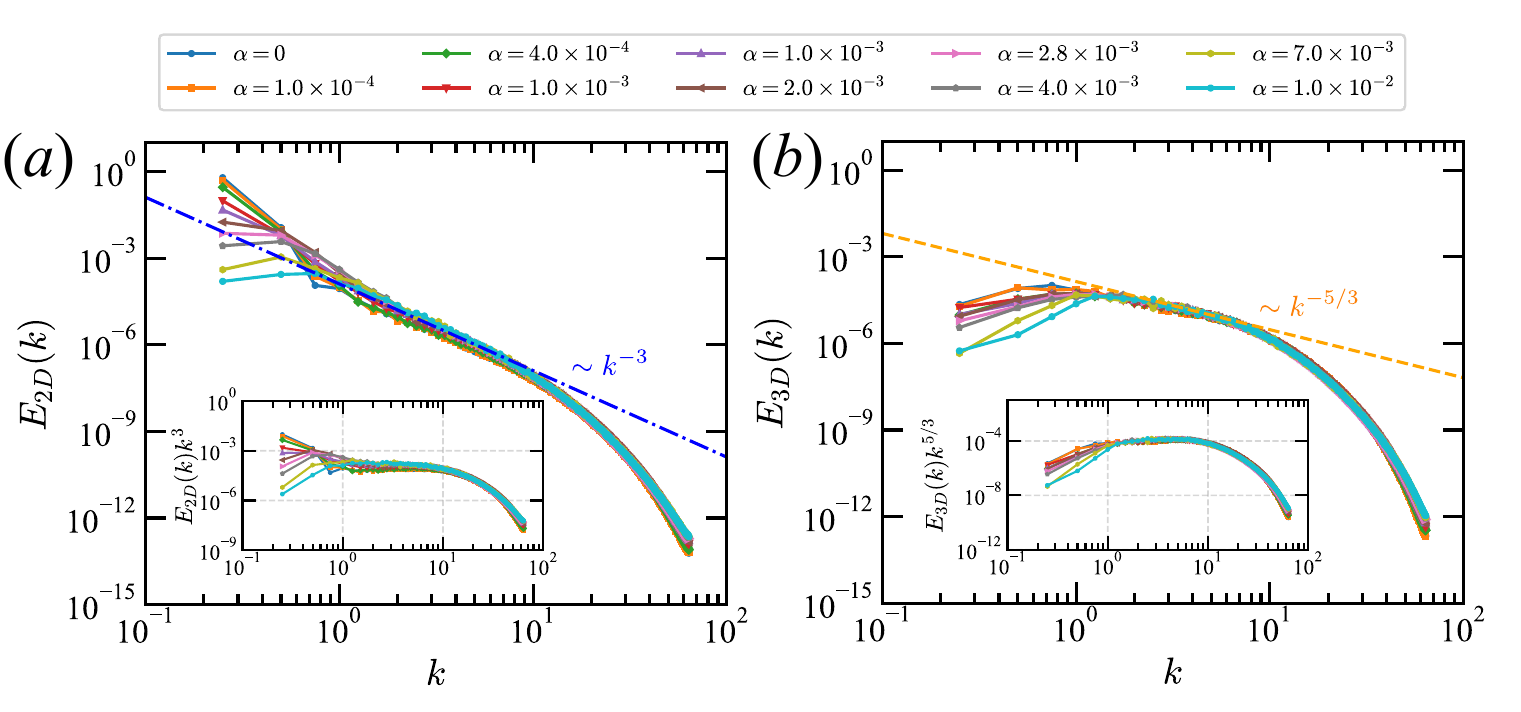}}
	\caption{Energy spectrum for (a) the 2D manifold $E_{2D}(k)$, and (b) the 3D manifold $E_{3D}(k)$ for all cases. The orange dashed line denotes $E(k)\sim k^{-5/3}$, and the blue dot-dashed line denotes $E(k)\sim k^{-3}$. The insets correspond to compensated plots of $E_{2D}(k)$ and $E_{3D}(k)$.}
	\label{fig:power_spectrum}
\end{figure}

Using the above decomposition, we compute the energy spectra of the 2D and 3D manifolds, $E_{2D}(k)$ and $E_{3D}(k)$, along with that of the full velocity field, $E(k)$. Figure \ref{fig:power_spectrum_single}(a) shows these spectra for $\alpha=1\times10^{-3}$ as a representative example. In the range $10^0\lesssim k\lesssim 10^1$, the 2D manifold exhibits the scaling $E_{2D}(k)\sim k^{-3}$, while the 3D manifold follows $E_{3D}(k)\sim k^{-5/3}$, consistent with previous studies on LSVs \citep{rubio_2014_prl}. The $k^{-3}$ scaling of $E_{2D}$ is reminiscent of the KLB enstrophy cascade \citep{alexakis_2018_physrep}; however, the spectral slope alone cannot distinguish between a forward enstrophy cascade and an inverse energy cascade. To resolve this ambiguity, we compute the energy fluxes $\Pi_{3D}$ and $\Pi_{2D}$ using the filtering method \citep{alexakis_2018_physrep}, shown in figure \ref{fig:power_spectrum_single}(b). The negative $\Pi_{2D}$ observed over the same wavenumber range confirms that the 2D manifold undergoes a net upscale energy transfer, ruling out a forward enstrophy cascade. To further assess the robustness of these scalings, figures \ref{fig:power_spectrum}(a) and (b) show the spectra $E_{2D}$ and $E_{3D}$ for all cases. Similar $k^{-3}$ scaling for the 2D manifold and $k^{-5/3}$ scaling for the 3D manifold can be found from all cases. Since the linear friction becomes significant at mainly large scales, data with different $\alpha$ collapse together at small scales for both $E_{2D}$ and $E_{3D}$. These results confirm that the inverse cascade of the 2D manifold, when sustained by energy input from the 3D manifold, yields a $k^{-3}$ scaling rather than the classical $k^{-5/3}$. 

A direct consequence of the $k^{-3}$ scaling is a modified prediction for the $\alpha$-dependence of the LSV scale. Following the same physical argument as equation \eqref{equ:l_alpha_k53} but substituting $k^{-5/3}$ with $k^{-3}$, we obtain
\begin{equation}
\epsilon_{3D}\sim\alpha\int^{k_\alpha} k^{-3} dk\sim \alpha k_\alpha^{-2},
\end{equation}
where $\epsilon_{3D}$ is the total energy injection rate from the 3D manifold related to the upscale energy transfer process. It further yields
\begin{equation}
	\label{equ:l_alpha_k3}
	k_\alpha\sim\alpha^{1/2},\text{ and }L_\alpha\sim\alpha^{-1/2}.
\end{equation}
This prediction is in excellent agreement with the observed LSV radius in figure \ref{fig:rlsv}(b). Equation \eqref{equ:l_alpha_k3} thereby resolves the discrepancy between the $\alpha^{-3/2}$ scaling predicted by the KLB theory (using $E(k)\sim k^{-5/3}$) and the observed $\alpha^{-1/2}$ scaling: the key distinction lies in the $k^{-3}$ spectrum of the 2D manifold rather than the classical $k^{-5/3}$. A key remaining question is why the 2D manifold exhibits a $k^{-3}$ energy spectrum. Although this scaling has been reported in previous studies, its underlying physical mechanism remains unclear.

\begin{figure}
	\centerline{\includegraphics[width=\textwidth]{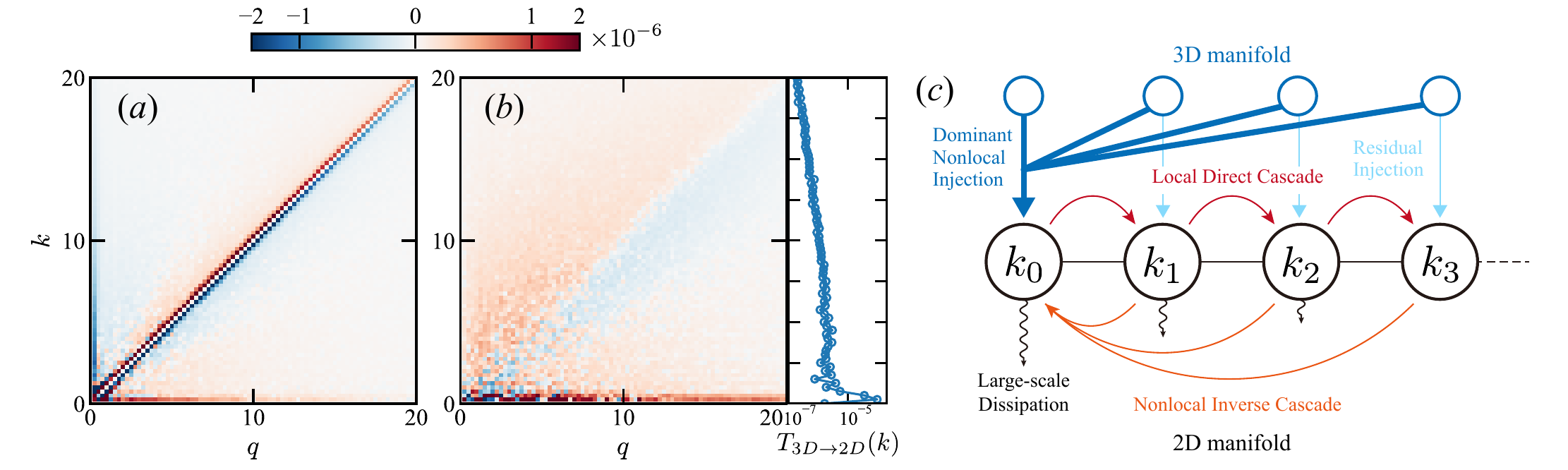}}
	\caption{Shell-to-shell energy flux (a) $T_{2D}(k,q)$ and (b) $T_{3D\rightarrow2D}(k,q)$ for $\alpha=1\times10^{-3}$. The blue circles in (b) refer to $T_{3D\rightarrow 2D}(k)\equiv\sum_qT_{3D\rightarrow2D}(k,q)$, which quantify the net energy injection to the 2D mode at wavenumber $k$. (c) An illustration for the energy pathways of the 2D manifold.}
	\label{fig:energy_pathway}
\end{figure}

A hint can be found by examining the detailed energy pathway of the 2D manifold. The 3D manifold spectrum $E_{3D}\sim k^{-5/3}$ is consistent with the Kolmogorov scaling for the direct inertial cascade. By contrast, the $k^{-3}$ scaling of the 2D manifold deviates from the KLB prediction $E(k)\sim k^{-5/3}$ for a local inertial inverse cascade, suggesting that the inverse cascade in the 2D manifold is not a local process. This is consistent with previous findings that the inverse cascade in rotating thermal convection is nonlocal \citep{rubio_2014_prl,aguirre_2020_prl}. To characterise the detailed energy pathway, we compute the shell-to-shell energy transfer rate \citep{mininni_2005_pre,mininni_2009_pof}. 

Figures \ref{fig:energy_pathway}(a) and (b) respectively show the shell-to-shell energy transfer rate within the 2D manifold, $T_{2D}(k,q)$,

\begin{equation}
	\label{equ:t_kq_2D}
	T_{2D}(k,q)\equiv-\langle \mathbf{u}_{2D}^k\cdot\left[(\mathbf{u}_{2D}\cdot\nabla)\mathbf{u}_{2D}^q\right]\rangle_V,
	\end{equation}
and the 3D-to-2D energy transfer rate $T_{3D\rightarrow 2D}(k,q)$,

\begin{equation}
	\label{equ:t_kq_3D}
	T_{3D\rightarrow2D}(k,q)\equiv-\langle \mathbf{u}_{2D}^k\cdot\overline{\left[(\mathbf{u}_{3D}\cdot\nabla)\mathbf{u}_{3D}^q\right]}\rangle_V,
	\end{equation}
where the superscript $\mathbf{u}^k$ denotes the bandpass filtered velocity for wavenumber between $k$ and $k+\delta k$. Here, $T_{2D}(k,q)$ quantifies the energy transferred from wavenumber $q$ to $k$ within the 2D manifold, while $T_{3D\rightarrow2D}(k,q)$ describes the energy injected from wavenumber $q$ of the 3D manifold into wavenumber $k$ of the 2D manifold. Previous numerical studies have shown that both transfers exhibit nonlocal behaviour \citep{rubio_2014_prl,aguirre_2020_prl,cai_2021_apj,de_wit_2022_jfm}. In figure \ref{fig:energy_pathway}(a), the smallest wavenumbers $k$ receive energy from a broad range of $q>k$, confirming a nonlocal inverse energy cascade. Simultaneously, energy is transferred along the diagonal ($k=q+\delta k$), corresponding to a local direct cascade. Although both direct and inverse transfers coexist, the net scale-to-scale flux is in the upscale direction, as confirmed by figure \ref{fig:power_spectrum_single}(b).

The 3D-to-2D energy injection also exhibits significant nonlocality. In figure \ref{fig:energy_pathway}(b), a dominant off-diagonal injection appears at the smallest wavenumbers, indicating that a broad range of wavenumbers $q$ in the 3D manifold directly inject energy into the largest scale of the 2D manifold, bypassing intermediate scales. This is consistent with previous studies \citep{rubio_2014_prl,aguirre_2020_prl}. The blue circles in figure \ref{fig:energy_pathway}(b) show $T_{3D\rightarrow2D}(k)\equiv\sum_qT_{3D\rightarrow2D}(k,q)$, the net energy injection into mode $k$ of the 2D manifold from the 3D manifold. A clear peak at $k\approx0.25$ marks this dominant large-scale injection. This dominant large-scale injection is then redistributed to smaller scales through the local direct cascade, as indicated by the diagonal of figure \ref{fig:energy_pathway}(a). 

An worth-mentioning observation is that the residual injection at higher wavenumbers is not negligible. As shown in figure \ref{fig:power_spectrum_single}(b), the inverse cascade transfers energy upscale starting from $k\gtrsim 10^0$. Since, by equation \eqref{equ:2d_manifold}, the only energy source for the 2D manifold is the 3D-to-2D injection, sustaining an inverse cascade at $k\gtrsim 10^0$ requires energy input at those wavenumbers. It follows that the dominant large-scale injection does not directly drive the inverse cascade; instead, it is the residual injection at $k\gtrsim 10^0$ that actually maintains it, visible as the off-peak values of $T_{3D\rightarrow2D}(k)$ in figure \ref{fig:energy_pathway}(b).

Figure \ref{fig:energy_pathway}(c) summarises the energy pathways of the 2D manifold. Based on the discussion above, we assume that the dominant nonlocal injection at $k_0$ is transferred to small scales by the local direct cascade and does not directly participate in the inverse cascade. The two key mechanisms maintaining the inverse cascade and the formation of the LSV are therefore the residual 3D-to-2D injection at intermediate wavenumbers and the nonlocal upscale energy transport. The nonlocality of the inverse cascade implies a direct coupling between the LSV scale and a broad range of smaller scales. Building on this picture, we develop a theoretical explanation for the $k^{-3}$ scaling of the 2D manifold in the next section.

\subsection{Scale-invariance for the 2D manifold}
\label{sec:scale_invariance}

\begin{figure}
	\centerline{\includegraphics[width=\textwidth]{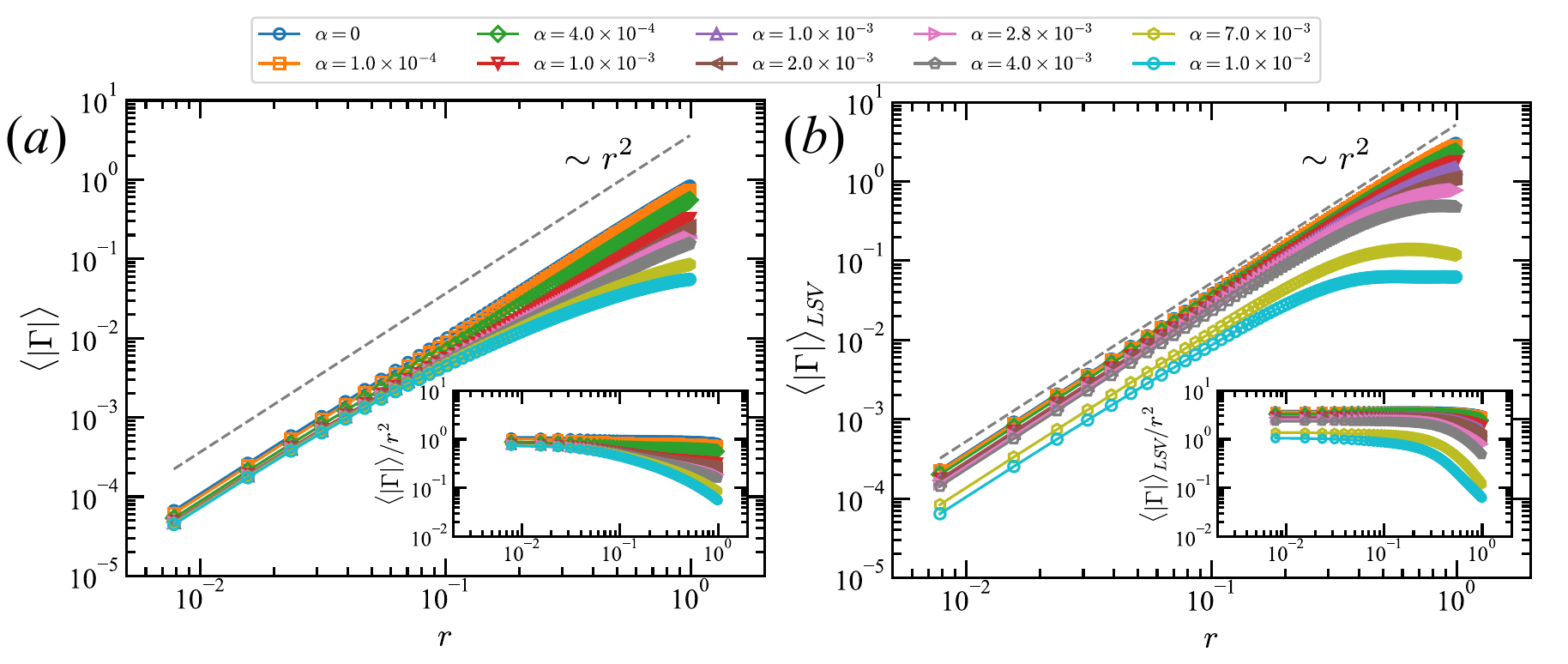}}
	\caption{Mean absolute circulation $\langle|\Gamma|\rangle$ as a function of $r$, averaged over (a) the whole 2D manifold, and (b) the LSV region. The insets present the normalized circulation $\langle|\Gamma|\rangle/r^2$.}
	\label{fig:circulation}
\end{figure}

As discussed above, the LSV scale couples with the various smaller scales through a nonlocal inverse cascade. This nonlocal coupling suggests that the dynamics at a certain small scale $r_n$ is not solely determined by local interactions among neighbouring scales. Instead, the LSV acts as a large-scale background flow that continuously distorts and reorganizes smaller-scale motions. We assume that such reorganization process by the LSV operates through a shear strain rate $S_{LSV}$ at the LSV scale

\begin{equation}
	S_{LSV}\sim U_{LSV}/r_{LSV}\sim U_{LSV}k_{LSV}.
\end{equation}
On the other hand, the local shear strain rate for a small scale $k_n$ is given by

\begin{equation}
	S_{n}\sim u(r_n)/r_n\sim u(k_n)k_n.
\end{equation}
Here $r_{LSV}$ and $r_n$ are respectively the LSV length scale and the local length scale corresponding to $k_n$. The LSV shear strain $S_{LSV}$ acts directly on small-scale vortices at $k_n$, competing with their local shear strain $S_n$. 

If $S_n<S_{LSV}$, the local strain associated with a local vortex is too weak compared with the LSV-induced shearing, meaning that the background LSV shear distorts the local flow faster than its self deformation.
The small scale vortex should then be stretched, tilted, and reorganized by the large-scale flow. On the other hand, if $S_n>S_{LSV}$, the small-scale vortices are too energetic for the LSV to control, thus the connection withe LSV scale could be lost then. The marginal condition is therefore the balance between $S_n$ and $S_{LSV}$, i.e.

\begin{equation}
\label{equ:strain_balance}
k_n u(k_n)\sim k_{LSV}U_{LSV}.
\end{equation}
Since both $k_{LSV}$ and $U_{LSV}$ are constant once the system reaches a statistically steady state, equation \eqref{equ:strain_balance} implies that the product $k_n u(k_n)$ is the same for all scales $k_n$ — i.e. the shear strain rate is scale-invariant. Equivalently, the coarse-grained vorticity $\xi(r)\equiv u(r)/r$ is scale-invariant. Since $\xi(r)$ is predicted to be constant across scales, a direct observational test is provided by the mean circulation. The circulation over a loop of radius $r$ scales as

\begin{equation}
	\label{equ:invariance_circulation}
	\Gamma(r)\equiv\oint_{|x|=r} \boldsymbol{u}\cdot \mathrm{d}\boldsymbol{l}
	=\int_{A(r)} \xi\,\mathrm{d}A
	\sim \xi(r)r^2,
\end{equation}
where $A(r)$ is the area enclosed by the loop. Since $\xi(r)$ is scale-invariant, equation \eqref{equ:invariance_circulation} gives $\langle|\Gamma(r)|\rangle\sim r^2$.

Figure \ref{fig:circulation} presents $\langle|\Gamma(r)|\rangle$ for the 2D manifold, averaged over (a) the whole system, and (b) the LSV region. Both statistics exhibit a clear $r^2$ scaling for small $\alpha$. For large $\alpha$, the LSV occupies a relatively small fraction of the domain. In this case, the spatial average over the whole system is significantly affected by the background turbulence, and may deviate from the $r^2$ scaling when the loop exceeds the LSV region. In contrast, when the statistics are restricted to the LSV region, a clear $r^2$ scaling is observed for all $\alpha$. The insets show the normalized circulation $\langle|\Gamma|\rangle/r^2$, which can be interpreted as a coarse-grained vorticity at scale $r$. A constant $\langle|\Gamma|\rangle/r^2$ can be observed over an intermediate range of $r$, meaning that the coarse-grained vorticity is scale-invariant. This plateau ends when $r$ becomes comparable to the LSV size. As the LSV size decreases with increasing $\alpha$, the scale-invariant range correspondingly shifts to smaller $r$.

The strain-rate balance \eqref{equ:strain_balance}, and equivalently the scale-invariance of coarse-grained vorticity, directly implies $k u(k)\sim \xi(k)\sim k^0$, i.e. $u(k)\sim k^{-1}$. The 2D-manifold energy spectrum then follows

\begin{equation}
	\label{equ:ek_kminus3}
	E_{2D}(k)\sim \frac{u(k)^2}{k}\sim k^{-3}.
\end{equation}
We emphasize that a $k^{-3}$ spectrum can also arise in the classical 2D enstrophy cascade; here, the circulation evidence supports an alternative interpretation in which the observed $k^{-3}$ scaling reflects an LSV-dominated range with approximately constant coarse-grained vorticity. This provides a physical explanation for the spectrum in figure \ref{fig:power_spectrum}, and further leads to the $\alpha$-dependence of $R_{LSV}$.

Finally, we wish to note that the observed scaling $\langle|\Gamma(r)|\rangle \sim r^2$ at very small $r$ has a different origin from that at larger scales. In the limit $r \to 0$, the velocity field is smooth and can be expanded linearly, $\mathbf{u}(\mathbf{x}+\mathbf{r}) \approx \mathbf{u}(\mathbf{x}) + (\mathbf{r}\cdot\nabla)\mathbf{u}$, which leads trivially to $\Gamma(r) \sim \xi r^2$. This small-scale $r^2$ scaling therefore reflects local smoothness of the flow and does not carry dynamical information about the cascade. In contrast, the $r^2$ scaling observed in the large scales, where the flow is turbulent and influenced by the inverse cascade, indicates that the coarse-grained vorticity $\xi(r) \sim |\Gamma(r)|/r^2$ becomes approximately scale invariant. It is this nontrivial, finite-range plateau in $\langle|\Gamma|\rangle/r^2$ that supports the interpretation of a strain-balanced, nonlocally coupled regime underlying the $k^{-3}$ spectrum.

\section{Conclusion}
\label{sec:conclusion}

We use direct numerical simulation of rapidly rotating Rayleigh--B\'enard convection to investigate how a linear friction with coefficient $\alpha$ controls the characteristic size of large-scale vortices (LSVs). In this study, we change $\alpha$ over a range $10^{-4}$--$10^{-2}$ and examine the scale and energy transfer of the LSVs. 

We observe a clear and robust $\alpha$-dependence of the LSV size, which can be quantified by both $R_m$ (a fitting parameter, corresponding to the radius of maximum azimuthal velocity) and $R_{LSV}$ (corresponding to the radius of zero vorticity). When friction is absent or weak, the LSV size is constrained by the finite domain and becomes essentially independent of $\alpha$. Once friction becomes significant, both $R_m$ and $R_{LSV}$ follow a scaling $\alpha^{-1/2}$. This differs noticeably from the prediction $L_{\alpha}\sim \alpha^{-3/2}$ based on the KLB inverse cascade spectrum $E(k)\sim k^{-5/3}$. 

To explain this discrepancy, we decomposed the flow into the vertically averaged 2D manifold (barotropic/slow mode) and the remaining 3D manifold (baroclinic/fast modes). We find that, over the range of wavenumbers associated with the LSV-building upscale transfer, the 2D manifold spectrum follows a robust $E_{2D}(k)\sim k^{-3}$, while the 3D manifold exhibits a $k^{-5/3}$ scaling, consistent with the direct inertial cascade. The energy flux of the 2D manifold in the same range is negative (upscale), indicating that the observed $k^{-3}$ is not simply the classical KLB forward enstrophy-cascade spectrum. Based on the observed $k^{-3}$ spectrum, one can directly obtain $k_\alpha\sim \alpha^{1/2}$ and $L_\alpha\sim \alpha^{-1/2}$ according to the standard friction-cutoff argument, which is consistent with the observed scaling for $R_m$ and $R_{LSV}$. Moreover, our results demonstrate that the absence of system-scale vortex can still yield a $k^{-3}$ spectrum, which is consistent with the experiment by \citet{zhu_2024_jfm}. The remaining question is the physical interpretation of the spectrum $k^{-3}$.

Such $k^{-3}$ scaling for the inverse energy cascade has been reported in many other systems with energy condensation, such as 2D turbulence \citep{chertkov_2007_prl}, rotating turbulence \citep{sen_2012_pre}, and thin-layer turbulence \citep{zhu_2024_jfm}. Yet, the physical interpretation for this scaling relationship remains unclear. Early simulations of forced 2D turbulence already suggested that coherent vortices can affect spectral statistics \citep{chertkov_2007_prl,legras_elepl_1988_hneftt}, thus we can expect the deviation with the classic 2D turbulence is attributed to the formation of LSV. Our interpretation the begins from depicting energy pathway for the LSV formation. Shell-to-shell energy analysis shows that both the energy transfer within the 2D manifold and the energy injection from the 3D manifold are strongly nonlocal. A dominant injection into the largest scales is identified, consistent with previous studies \citep{de_wit_2022_jfm,aguirre_2020_prl}. However, sustaining the inverse energy cascade also requires that residual injection across a broad range of intermediate scales is non-negligible. Additionally, the inverse energy transfer within the 2D manifold exhibits significant nonlocality, meaning that the LSV scale absorbs energy directly from various smaller scales.

The nonlocal coupling between the LSV scale and a broad range of smaller scales suggests a marginal balance between the LSV shear strain rate $S_{LSV}$ and the local shear strain rate $S_n$ at scale $k_n$. This strain-rate balance imples a scale-invariant coarse-grained vorticity $\xi(r)\equiv u(r)/r$, which directly yields $u(k)\sim k^{-1}$ and therefore $E_{2D}(k)\sim k^{-3}$. The prediction is verified by the mean circulation $\langle|\Gamma(r)|\rangle$, which exhibits a scaling $\langle|\Gamma(r)|\rangle\sim r^2$, confirming that the coarse-grained vorticity $\langle|\Gamma|\rangle/r^2$ is indeed scale-invariant.

In this paper, we propose a new scaling relationship for the friction-controlled cutoff length scale. The key insight is that the inverse energy cascade in the presence of an LSV does not follow the KLB spectrum $E(k)\sim k^{-5/3}$, but instead exhibits a $k^{-3}$ spectrum. A physical explanation for this $k^{-3}$ spectrum is further provided, basetad on the scale-invariance of the coarse-grained vorticity. To test the generality of this argument, we have conducted simulations of 2D forced turbulence with energy condensation. The results are shown in Appendix \ref{app:forced-2d}. In this system, similar energy spectral and mean circulation scalings are observed. Such similarity suggests the potential for generalising the interpretation of $E(k)\sim k^{-3}$ to other turbulence systems, and extending the argument to further model systems or astrophysical data would be worth pursuing in future work. 

These findings also have broader implications for geophysical and astrophysical flows, where large-scale vortices arising from rotating turbulent environments widely exist. The Ekman friction shares a similar mathematical form with the large-scale friction considered in this study. The present results therefore suggest that using an assumed $k^{-5/3}$ inverse-cascade spectrum to infer an $\alpha^{-3/2}$ cutoff length scale may significantly misestimate the size of large-scale vortices when the barotropic dynamics instead exhibit a $k^{-3}$ spectrum. Such differences can lead to substantial deviation when extrapolating laboratory or numerical results to planetary and stellar regimes. It is interesting to extend the current theoretical framework to these systems.

\begin{bmhead}[Acknowledgements.]
	G.Y.D., T.Y.P., H.Y.Z. and K.Q.X. thank the discussion with Prof. Roberto Benzi, Prof. Wanying Kang and Prof. Jin-Qiang Zhong, and acknowledge the support from the Center for Computational Science and Engineering of Southern University of Science and Technology.
\end{bmhead}
\begin{bmhead}[Funding.]
	This study is supported by the funds from the National Natural Science Foundation (NSFC) (grant No.12302282, 12594530010, 12595302), and the Startup Foundation of Fudan University.
\end{bmhead}
\begin{bmhead}[Declaration of interests.]
	The authors report no conflict of interest.
\end{bmhead}

\appendix
\section{LSV identification and data information}
\begin{figure}
	\centerline{\includegraphics[width=\textwidth]{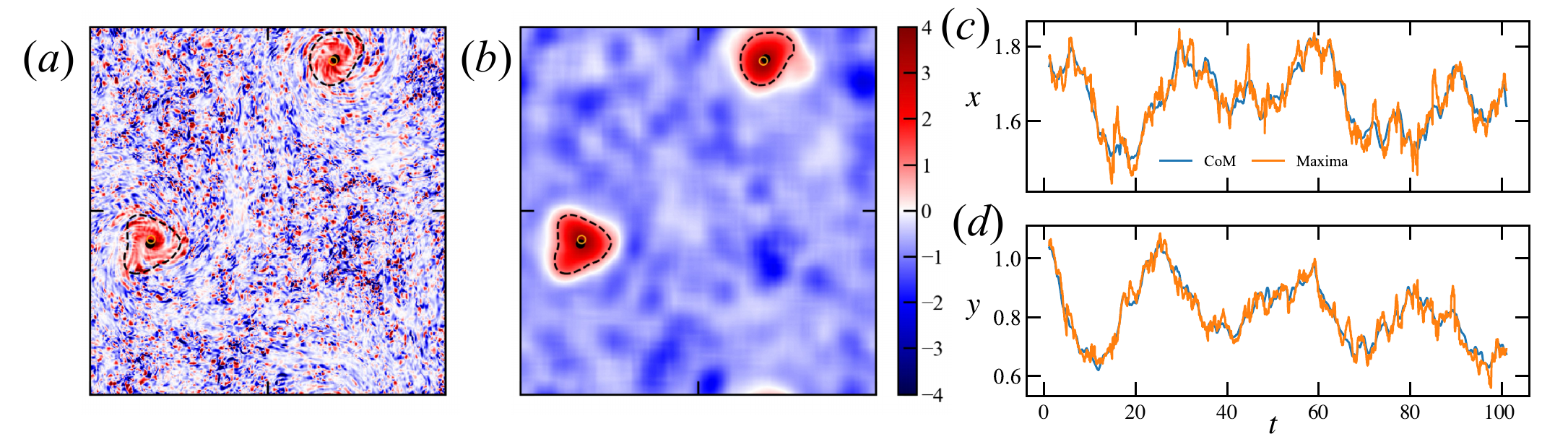}}
	\caption{(a) Distribution of the original Q-field $Q/Q_{std}$, (b) the filtered Q-field $Q_f/Q_{f,std}$. The black dashed curves denote the area of LSVs defined by the contour $Q_f/Q_{f,std}=1$, the black close circle and orange open circle respectively refer to the center of LSV obtained from the CoM and Maxima strategies. The time trace for the position of a LSV in x- and y-coordinates are shown in (c) and (d), respectively. The blue and orange curves respectively refer to the CoM and Maxima strategies.}
	\label{fig:identification}
\end{figure}

Our LSV-identification procedure is based on methods previously used for convective Taylor columns (CTCs), with modifications tailored to the strongly turbulent background. CTCs are another class of vortices extensively studied in rotating Rayleigh--B\'{e}nard convection; previous studies often use the $Q$-criterion to identify CTCs \citep{chong_2020_sa,ding_2021_ncomms}. The $Q$ value is defined as
\begin{equation}
	Q\equiv\frac{1}{2}\left(\Omega^2-S^2\right),
\end{equation}
where $\Omega$ and $S$ are, respectively, the vorticity and strain-rate tensors. For CTCs, the vortex region is typically defined by $Q>Q_{c}$, where $Q_c$ is a threshold (often taken as the standard deviation of the $Q$ field). The CTC centre can then be defined as the location of the maximum $Q$ within this region. In the present system, the LSV coexists with strong background turbulence, which leads to substantial fluctuations in the $Q$ field. As demonstrated in figure \ref{fig:identification}(a), the LSV can still be seen in the raw $Q$ field, but obtaining a continuous region requires additional filtering. We therefore filter the $Q$ field using a Gaussian kernel,
\begin{equation}
	G(r)=\frac{6}{\pi\Delta^2}\exp\left(-\frac{6r^2}{\Delta^2}\right),
\end{equation}
where $\Delta$ is the characteristic filter width, chosen based on an estimate of the LSV size from vorticity snapshots. The filtered field is defined as $Q_f(r)=Q(r)*G(r)$, where $*$ denotes convolution. Figure \ref{fig:identification}(b) shows the filtered field $Q_f/Q_{f,std}$, where $Q_{f,std}$ is the standard deviation of $Q_f$. The filtering strongly reduces the influence of background turbulence. The black dashed curves denote the LSV region defined by the contour $Q_f/Q_{f,std}=1$, which yields a continuous LSV area. For the LSV centre, we find that using the maximum of $Q_f$ is noisy: residual turbulent fluctuations within the LSV can still shift the exact location of the maximum. We therefore define the LSV centre using the centre-of-mass (CoM) of the LSV region, with
\begin{equation}
	\vec{r}_{c}\equiv \frac{\int_{LSV} \vec{r} Q_f(r)\,\mathrm{d}A}{\int_{LSV} Q_f(r)\,\mathrm{d}A},
\end{equation}
where $LSV$ denotes the LSV region. The LSV centre is then defined as the CoM position. Time traces of the LSV position in the $x$ and $y$ directions are shown in figure \ref{fig:identification}(c) and (d), respectively, where the blue and orange curves correspond to the CoM and maximum-based definitions. The CoM strategy is markedly more stable: the maximum-based definition overestimates short-time wandering due to residual turbulent fluctuations, whereas the CoM better reflects the dynamics of an inertial structure. We therefore use the CoM strategy to define the LSV centre in the following analysis.

\begin{table}
	\begin{tabular*}{\textwidth}{@{\extracolsep{\fill}}cccccccccc@{}}
		Ra & Pr & Ek & $\Gamma$ & $\alpha$ & $R_{\mathrm{LSV}}$ & $\mu$ & $R_m$ & $L_0$ & $t_{\mathrm{stat}}$ \\
		\midrule
		$1\times10^8$ & 1 & $3\times10^{-5}$ & 4 & $0$ & 1.18 & 1.08 & 0.70 & 2.69 & 2000 \\
		$1\times10^8$ & 1 & $3\times10^{-5}$ & 4 & $1.0\times10^{-4}$ & 1.19 & 1.01 & 0.68 & 2.68 & 2000 \\
		$1\times10^8$ & 1 & $3\times10^{-5}$ & 4 & $4.0\times10^{-4}$ & 1.17 & 0.96 & 0.62 & 2.66 & 2000 \\
		$1\times10^8$ & 1 & $3\times10^{-5}$ & 4 & $1.0\times10^{-3}$ & 1.04 & 0.91 & 0.51 & 2.61 & 2000 \\
		$1\times10^8$ & 1 & $3\times10^{-5}$ & 4 & $1.4\times10^{-3}$ & 0.92 & 0.98 & 0.44 & 2.55 & 2000 \\
		$1\times10^8$ & 1 & $3\times10^{-5}$ & 4 & $2.0\times10^{-3}$ & 0.73 & 1.00 & 0.35 & 2.40 & 1000 \\
		$1\times10^8$ & 1 & $3\times10^{-5}$ & 4 & $2.8\times10^{-3}$ & 0.61 & 1.02 & 0.26 & 2.29 & 1000 \\
		$1\times10^8$ & 1 & $3\times10^{-5}$ & 4 & $4.0\times10^{-3}$ & 0.47 & 1.09 & 0.22 & 1.88 & 1000 \\
		$1\times10^8$ & 1 & $3\times10^{-5}$ & 4 & $7.0\times10^{-3}$ & 0.34 & 1.38 & 0.17 & 1.57 & 1000 \\
		$1\times10^8$ & 1 & $3\times10^{-5}$ & 4 & $1.0\times10^{-2}$ & 0.24 & 1.82 & 0.15 & 1.19 & 1000 
	\end{tabular*}
    \caption{Data information.}
    \label{tab:data_info}
\end{table}

\section{Forced two-dimensional turbulence}
\label{app:forced-2d}

To test whether the connection between the $k^{-3}$ spectrum and the $\langle |\Gamma(l)|\rangle \sim l^2$ circulation scaling has a general significancy for turbulence or is just a coincidence in the convection system, we perform simulations of two-dimensional forced turbulence in a doubly periodic box of size $(2\pi)^2$ with a resolution $N^2 = 512^2$, using a pseudospectral method with two-thirds dealiasing and fourth-order Runge--Kutta time stepping \citep{mininni_pc_2011_hmossp}. Energy is injected with random phases and amplitudes in a narrow spectral band centred at $k_f=16$ with width $\Delta k_f=1$. The energy injection rate $\varepsilon$ is normalized so that it is statistically being unity. We use standard Laplacian viscosity with $\nu=2\times10^{-3}$. To study condensation with LSVs in two-dimensional turbulence \citep{chertkov_2007_prl,xia_2008_prl}, we compare a frictionless case, $\alpha=0$, with a moderately linearly damped case, $\alpha=10^{-2}$. Following \citet{vankan_x_2025_ttcbd}, we consider the enstrophy-dissipation scale $\eta_{\omega}=\varepsilon^{-1/6}k_f^{-1/3}\nu^{1/2}$ and verify that $k_{\max}\eta_{\omega}\gg1$, with $k_{\max}=N/3$. 

\begin{figure}
\centering
\includegraphics[width=\textwidth]{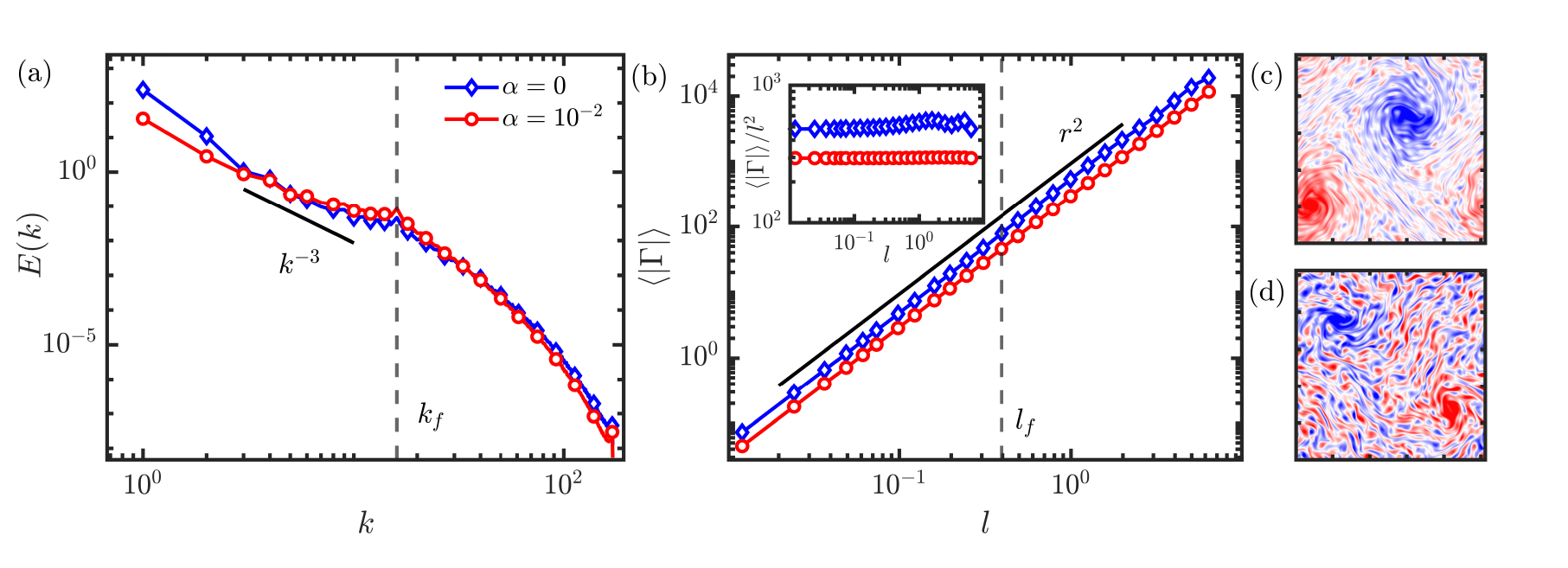}
\caption{Forced two-dimensional turbulence. (a) Energy spectra for the frictionless case $\alpha=0$ and the damped case $\alpha=10^{-2}$. The vertical dashed line marks the forcing wavenumber $k_f=16$ and the scaling $k^{-3}$ is given for reference. (b) Mean absolute circulation $\langle |\Gamma(l)|\rangle$ as a function of loop side length $l$. The vertical dashed line marks the forcing length $l_f$ and the scaling $l^2$ is given for reference. The inset presents the normalized circulation $\langle |\Gamma(l)|\rangle / l^2$ (c,d) Instantaneous vorticity fields for $\alpha=0$ and $\alpha=10^{-2}$, respectively.}
\label{fig:app_forced2D}
\end{figure}

Figure~\ref{fig:app_forced2D} shows that the same spectral and circulation signatures appear in condensed 2D turbulence (flow patterns are presented in figures~\ref{fig:app_forced2D}(c, d)). The energy spectra in figure~\ref{fig:app_forced2D}(a) show a large-scale range close to $E(k)\sim k^{-3}$ on the low-wavenumber side of the forcing wavenumber $k_f$. The mean absolute circulation over square loops $\langle |\Gamma(l)|\rangle$ in figure~\ref{fig:app_forced2D}(b) exhibits a clear $l^2$ scaling for both cases, with plateaus for $\langle |\Gamma(l)|\rangle / l^2$ found in the insert. These condensed two-dimensional turbulence results therefore suggest that the explanation given in section \ref{sec:scale_invariance} may extend to more general flow configurations dominated by large-scale coherent structures. 

\bibliographystyle{jfm}
\bibliography{reference}

\end{document}